\documentclass[11pt]{article}
\usepackage{amsmath}
\usepackage{amsfonts}
\usepackage{amssymb}
\usepackage{graphicx}
\usepackage{subfigure}
\usepackage{epsfig}

\def\bea{\begin{eqnarray}}
\def\eea{\end{eqnarray}}

\begin{document}
\begin{center}
\LARGE {\bf  Symmetry and quantum transport on networks
 }
\end{center}
\begin{center}
{\bf S. Salimi {\footnote {E-mail: shsalimi@uok.ac.ir}}, R. Radgohar
{\footnote {E-mail: r.radgohar@uok.ac.ir}},
M.M. Soltanzadeh {\footnote {E-mail: msoltanzadeh@uok.ac.ir}}}\\
 {\it Faculty of Science,  Department of Physics, University of Kurdistan, Pasdaran Ave., Sanandaj, Iran} \\
 \end{center}
\vskip 3cm
\begin{center}
{\bf{Abstract}}
\end{center}
We study the classical and quantum transport processes on some
finite networks and model them by  continuous-time random walks
(CTRW) and continuous-time quantum walks (CTQW), respectively. We
calculate the classical and quantum transition probabilities between
two nodes of the network. We numerically show that there is a high
probability to find the walker at the initial node for CTQWs on the
underlying networks due to the interference phenomenon, even for
long times. To get global information (independent of the starting
node) about the transport efficiency, we average the return
probability over all nodes of the network. We apply the decay rate
and the asymptotic value of the average of the return probability to
evaluate the transport efficiency. Our numerical results prove that
the existence of the symmetry in the underlying networks makes
quantum transport  be less efficient than the classical one. In
addition, we find that the increasing of the symmetry of these
networks decreases the efficiency of quantum transport on them.
\newpage
\section{Introduction}
Quantum walk(QW) as a generalization of random walk(RW) is obtained
by endowing the walker with quantum properties~\cite{ES, CCD1}. The
QW has been largely based on two standard variants, the discrete
time QW(DTQW)~\cite{DJ} and the continuous-time QW(CTQW)~\cite{ES}.
In recent years, the DTQWs have been investigated on
trees~\cite{CHKS},  random environment~\cite{NKQ},  single and
entangled particles~\cite{CMC, CSRLR, GAD}. The CTQWs have been
studied on line~\cite{GSR, jas1, jas3, konno}, cycle~\cite{SF,
xinp}, one-dimension regular network~\cite{XX50},  regular
graphs~\cite{jas2}, odd graphs \cite{shs}, cayley tree~\cite{MBO,SS,
konno2}, hypercube~\cite{KB}, small-world network~\cite{MPB}, star
graphs~\cite {salimi12, Xinpin}, dendrimer~\cite{OMV}, restricted
geometries~\cite{EAO}, Apollonian network~\cite{XL} and
one-dimensional and two-dimensional networks with periodic boundary
conditions~\cite{OAB,VMB}. Moreover, the decoherent QWs have been
considered on hypercube~\cite{GAAR}, cycle~\cite{VKBT, FST},
long-range interacting cycle~\cite{SRLS, SRLL} and one-dimension
regular network~\cite{SRR}.
\\Since the QWs may exploit quantum mechanical effects such as
entanglement and interference, the only special experimental
techniques can be candidate to implement them. Recently,  some
experimental implementations of both QW variants have been reported
e.g. on microwave cavities~\cite{SBC}, ground state atoms~\cite{WD},
the orbital angular momentum of photons~\cite{PZH}, waveguide
arrays~\cite{HBP} or Rydberg atoms~\cite{RC, M99}. \\Since RWs
generate the classical algorithms in computer science and model the
diffusion phenomena and non-deterministic motion on the complex
networks~\cite{MB}, it can be expected that their quantum
extensions(QWs) provide tool to implement quantum computing and to
model quantum processes~\cite{NCQ1}. In quantum computing, it is a
primary goal to determine when quantum computers can solve problems
faster than classical computers~\cite{BC3, CFG5}. Grover in 1996
proved that a classical computer requires $\mathcal{O}(N)$ steps to
finding marked item among $N$ items, whereas quantum computer can
solve it using only $\mathcal{O}(\sqrt{N})$ steps~\cite{LKG}.
Several implementations to Grover's algorithm have been reported by
CTQW~\cite{ES, CJG, AAO} and DTQW~\cite{AT, AAJKR, NSJKW}. To
implement Grover's algorithm, the all items must be accessible by
local moves. For instance, if items are located on the
one-dimensional line, traveling from one end of the line to the
other end  requires $N$ moves, thus the classical or quantum
algorithms can not find a marked item in less time than
$\mathcal{O}(N)$~\cite{CJG}. It is interesting to consider a
one-dimensional line and then gradually to change its structure so
that, in every step, items on it to be accessible by the lesser
moves, as shown in Figs. 1(a-e). \clearpage
\begin{figure}[h]
\vspace{4cm}\hspace{-5cm}\includegraphics{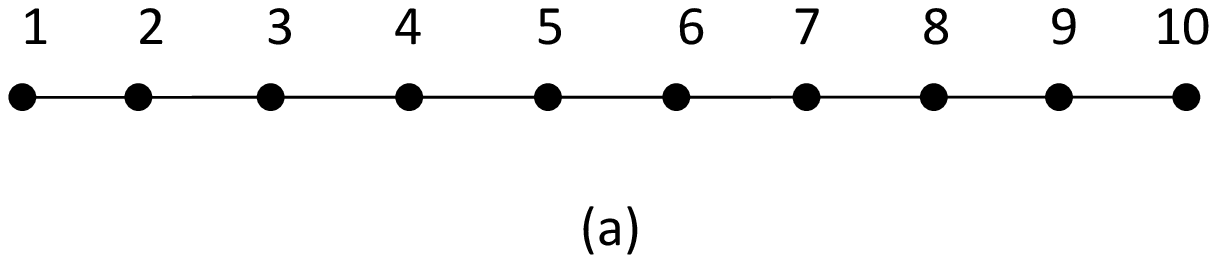}\vspace{-0.1cm}\hspace{9cm}
\includegraphics{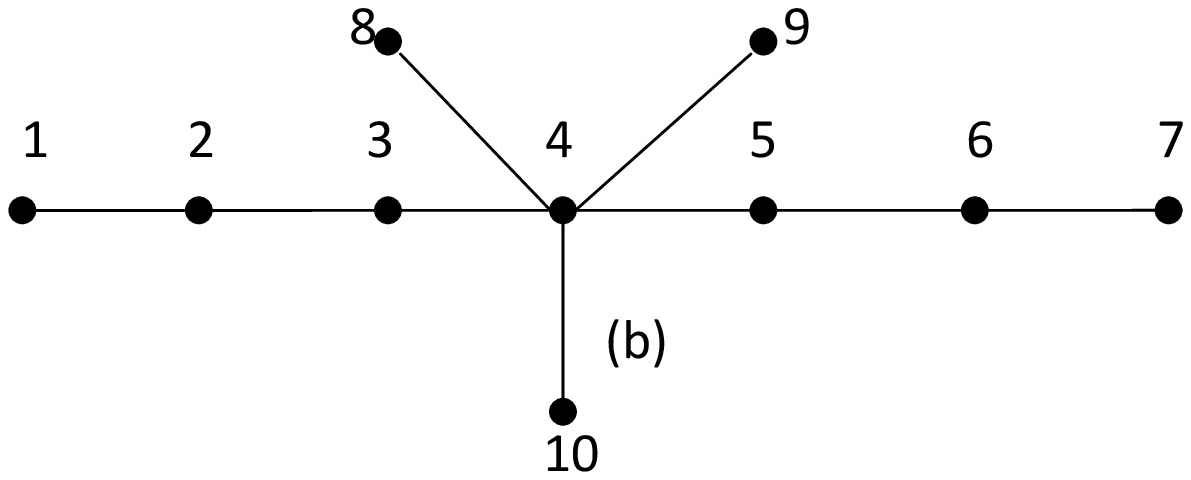}
\end{figure}

\begin{figure}[h]
\vspace{3.3cm}\hspace{-5cm}\includegraphics{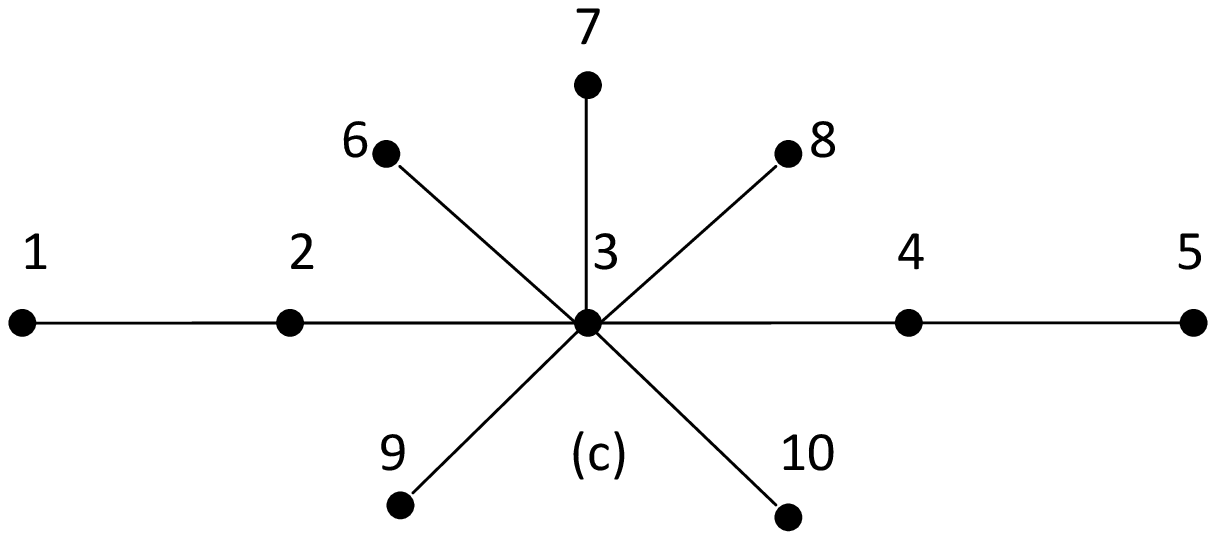}\vspace{0.1cm}\hspace{9cm}\includegraphics{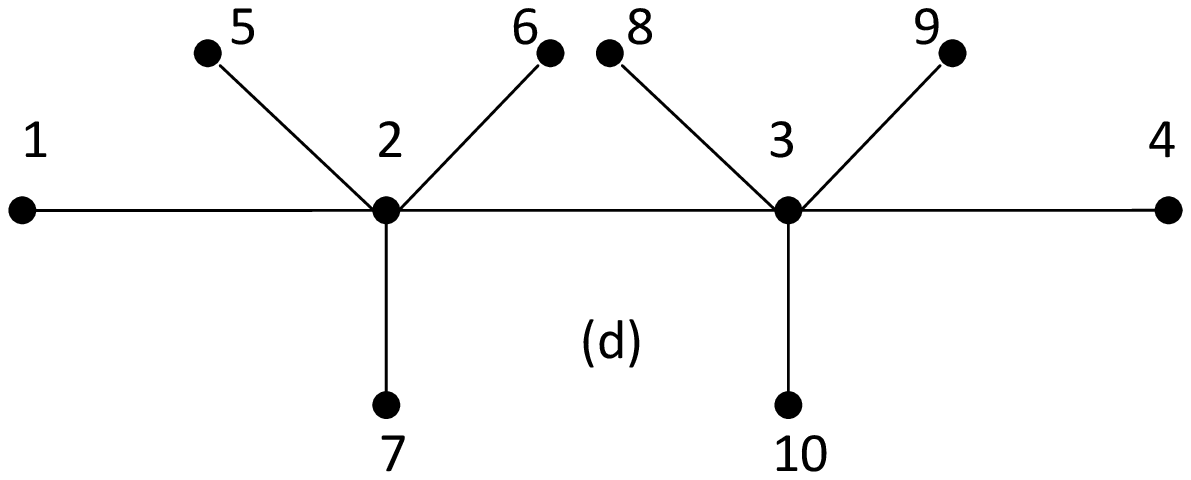}
\end{figure}

\begin{figure}[h]
\vspace{4.5cm}\hspace{0cm}\includegraphics{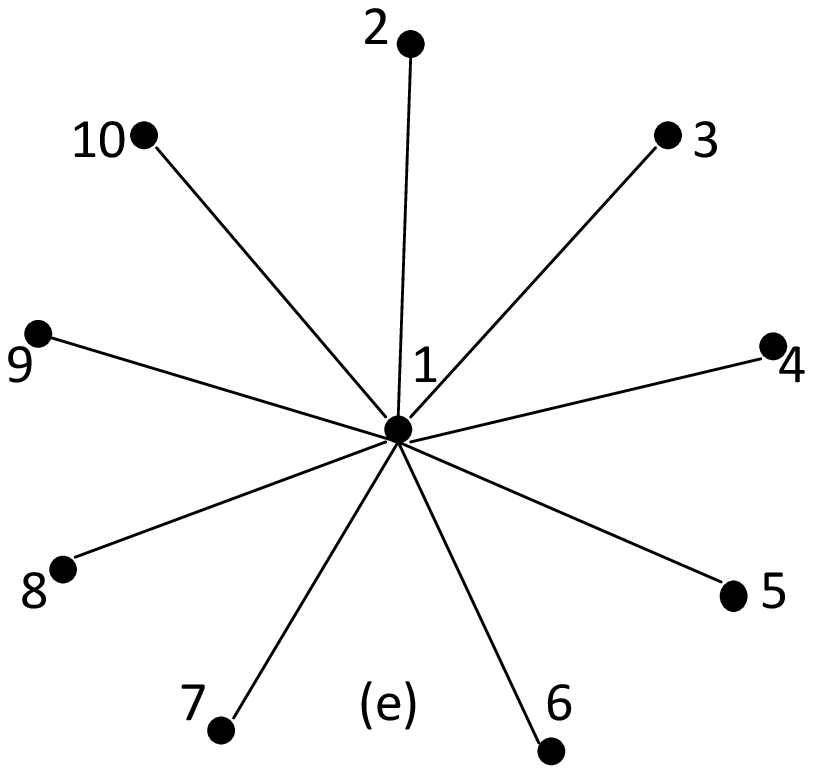}\caption{Figs. 1(a)-(e) show some networks
with ten nodes and nine edges.}
\end{figure}

On the other hand, in recent years the complex networks have been
applied to model very diverse systems in nature such as lattices in
solid state physics and condensed matter~\cite{RAALB, SNDJFM,
SBVLYM, RMJK}. In this modeling process, the components of system
and interactions among them are considered as nodes and bonds of
network, respectively. It is difficult to study the complex networks
having many nodes or bonds or both. But the most complex networks
are constructed from simpler networks which the number of nodes and
the number of bonds are smaller. Since these simpler networks can
successfully explain some features of whole network, there is an
upsurge of interest in studying of their various aspects. Hence, by
studying of RW and QW on the above mentioned simple networks, we
also highlight the interplay between the diffusion process and the
network symmetry. \\In this paper, we first calculate transition
probabilities between two nodes of network for the CTRWs and CTQWs,
then we numerically show that for CTQWs on the mentioned networks
due to interference phenomenon, even for long times, there is high
probability to find the walker at the initial node. Thus, the
quantum transition probability distribution on underlying networks
depends significantly on the initial node. To get a global
information about efficiency of walk, we average the return
probability over all nodes of network. Then, we evaluate the
transport efficiency on networks(a-e) by the rate decay and by the
asymptotic value of average return probability. The numerical
results show that the existence of symmetry in the mentioned
networks causes the quantum transport to be less efficient than the
classical counterpart. In other words, the results  denote that the
increasing of the symmetry of these networks decreases the
efficiency of quantum transport on them.
\\Our paper is structured as follows: after a brief summary of the
main concepts and of the formulae concerning CTQWs in Sec. 2, we
study the quantum transition probabilities and their long time
average on the mentioned networks, in Sec. 3. In Sec. 4, we
especially focus on the average return probabilities and the
efficiency of classical and quantum walks. Finally, in Sec. 5 the
conclusions are presented.

\section{CTQWs on networks}
In general, every network can be characterized by a graph $G$ that
consists of a finite nonempty set $V$ of $N$ vertices together with
a prescribed set $X$ of q unordered pairs of distinct vertices of
$V$ so that each pair $x=\{u,v\}$ of vertices in $X$ is a edge of
$G$~\cite{FH}.
\\Algebraically, a graph can be described by its adjacency matrix
$A$ whose elements are $A(i,j)=1$ for $(i,j)\in X$ and $A(i,j)=0$
otherwise. The Laplacian operator is then defined as $L=Z-A$, where
$Z$ is a diagonal matrix and $Z_{j,j}$ is the degree of vertex $j$.
Classically, the evolution of CTRW is governed by the following
master equation~\cite{GH}:

\begin{eqnarray}\label{1}
    \frac{d}{dt} P_{k,j}=\sum_{l=1}^{N} T_{kl}P_{l,j},
   \end{eqnarray}

where $P_{k,j}(t)$ is the conditional probability to find the walker
at time $t$ at node $k$ when starting at node $j$.
\\The matrix $T$ is the transfer matrix of the walk,
$T=(T_{kj})$, and relates to the Laplace matrix through $T=-\gamma
L$, where we assume an unbiased CTRW so that
the transmission rates of all bonds are equal and set $\gamma\equiv1$.\\
The CTQW as the quantum mechanical extension of CTRW is obtained by
replacing the Hamiltonian of the system by the classical transfer
matrix, $ H=-T$~\cite{ES}. \\We denote the state associated with
node $j$ of network as $|j\rangle$ and take the set $\{|j\rangle\}$
to be orthonormal. Thus, the solution of Eq. (1) is
 \begin{eqnarray}\label{2}
   P_{k,j}(t)=\langle k|e^{t T}|j\rangle.
\end{eqnarray}
Quantum mechanically, the states $|j\rangle$ span the whole
accessible Hilbert space. The time evolution of state $|j\rangle$
starting at time $t_{0}$ is given by
$|j,t\rangle=U(t,t_{0})|j\rangle$, where
$U(t,t_{0})=\exp[-iH(t-t_{0})]$ is the quantum mechanical evolution
operator. Hence, transition amplitude $\alpha_{k,j}(t)$ from state
$|j\rangle$ at the time 0 to state $|k\rangle$ at time $t$ is
\begin{eqnarray}\label{3}
 \alpha_{k,j}(t)=\langle k|e^{-iHt}|j\rangle.
\end{eqnarray}
From the Schrodinger Equation(SE), we obtain
\begin{eqnarray}\label{4}
i\frac{d}{dt}\alpha_{k,j}(t)=\sum_{l=1}^{N}H_{kl}\alpha_{l,j}(t),
\end{eqnarray} where $\hbar=1$.
We assume that $E_{n}$ and $|q_{n}\rangle$ denote the $n$th
eigenvalue and eigenvector of $H$, respectively. The classical and
quantum transition probabilities between two nodes can be written
as~\cite{OMBE},
\begin{eqnarray}\label{5}
P_{k,j}(t)=\sum_{n=1}^{N}e^{-tE_{n}}\langle k|q_{n}\rangle\langle
q_{n}|j\rangle
\end{eqnarray}

\begin{eqnarray}\label{6}
\pi_{k,j}(t)=|\alpha_{k,j}(t)|^{2}=|\sum_{n=1}^{N}e^{-itE_{n}}\langle
k|q_{n}\rangle\langle q_{n}|j\rangle|^{2}.
\end{eqnarray}
Note that the normalization condition for $P_{k,j}(t)$  is
$\sum_{k=1}^{N}P_{k,j}(t)=1$ and for $\alpha_{k,j}(t)$ is
$\sum_{k=1}^{N}|\alpha_{k,j}(t)|^{2}=1$.
\\Classically, the transition probabilities converge to the equip-partitioned probability $\frac{1}{N}$, whereas
the quantum probabilities do not reach any finite value but after
some time fluctuate about a constant value. This value is determined
by the long time average(LTA) which is defined by using Eq.
(6)~\cite{VMB}:
\begin{eqnarray}\label{12}
  \chi_{k,j} =\lim_{T\longrightarrow\infty}\frac{1}{T}\int^{T}_{0}\pi_{k,j}(t)dt =\displaystyle\sum_{n,m}\delta_{E_{n},E_{m}}\langle k|q_{n}\rangle\langle
   q_{n}|j\rangle\langle j|q_{m}\rangle\langle q_{m}|k\rangle,
\end{eqnarray}
where $\delta_{E_{n},E_{m}}=1$ for $E_{n}=E_{m}$ and
$\delta_{E_{n},E_{m}}=0$ otherwise. Since some eigenvalues of $H$
may be degenerate, the sum in the Eq. (7) contains terms belonging
to different eigenvectors.

\section{Quantum transition probabilities}
In this section, we illustrate the importance of choosing the
initial node in CTQWs process on the networks. For this aim, we
first consider the quantum transition probabilities in intermediate
times when the walker started from a special node. Then, we
generalize our result for  long times and the whole nodes by
studying the LTA probabilities.
\\To obtain the quantum transition
probabilities(see Eq. (6)), we need all the eigenvalues and
eigenvectors of $H$, thus we make use of standard software MATLAB.
\\In Fig. 2(a-e) we present the quantum transition probabilities $\pi_{k,j}(t)$ for
networks(a-e) where $j=5,4,...,1$, respectively.  Figs. 2(a-e) show
the high values for $\pi_{5,5}$, $\pi_{4,4}$, $\pi_{3,3}$,
$\pi_{2,2}$ and $\pi_{1,1}$. In other words, these figures  prove
that at short times(from $t=1s$ to $t=20s$), the quantum return
probability is large and thus the CTQWs hold such dependence
significantly on given starting
 node. \\Now it is natural to pose the following questions:
\\Whether the such behavior continues at long times?
\\Do the CTQWs hold  such dependence on the other starting nodes?
\\To address these questions, in Figs.
3(a-e) we plotted the LTA probabilities for all nodes $k$ and $j$
pertaining to networks(a-e), respectively. \\The axes $x$ and $y$
show nodes $k$ and $j$ of network respectively, and $\chi_{k,j}$ is
presented on the axis $z$. Figs. 3(a-e) show the symmetry
characterizing the quantum transition probability, namely
$\chi_{k,j}=\chi_{j,k}$. This can be derived directly from Eq. (7),
recalling that $H$  itself is symmetric and real.

\begin{figure}[h]
\vspace{6cm}\hspace{-4cm}\includegraphics{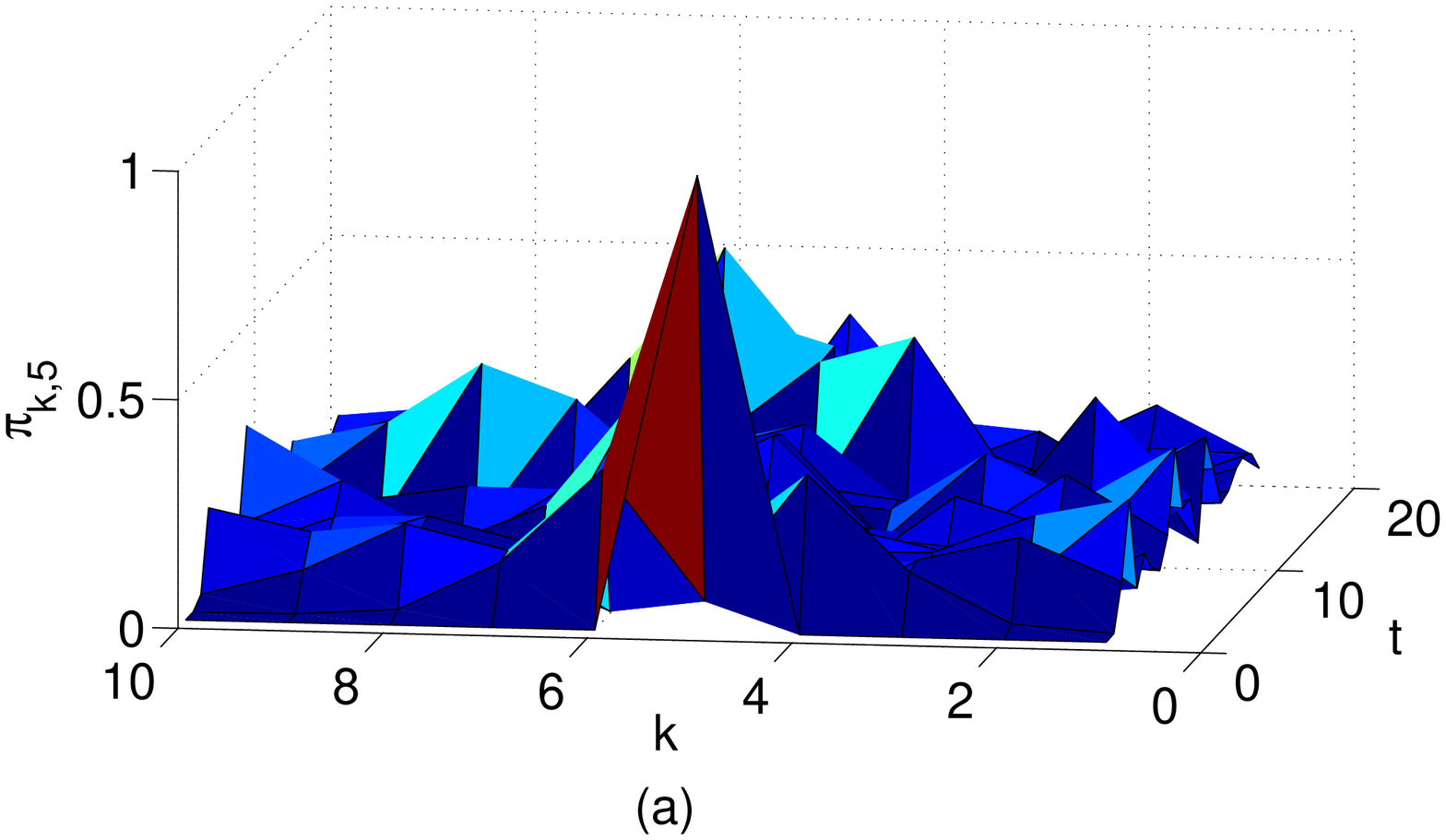}\vspace{-0.1cm}\hspace{10cm}
\includegraphics{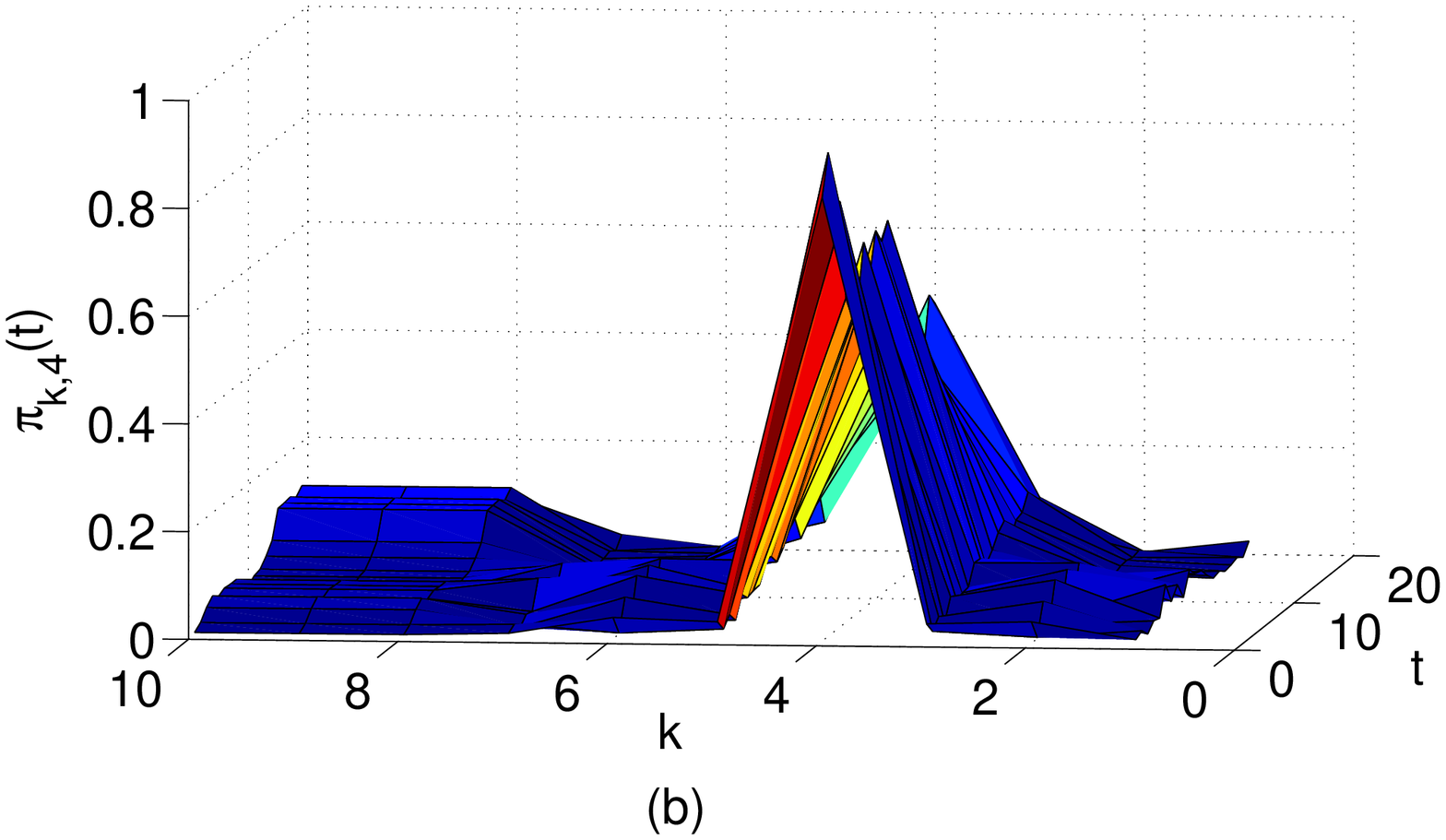}
\end{figure}

\begin{figure}[h]
\vspace{6cm}\hspace{-4cm}\includegraphics{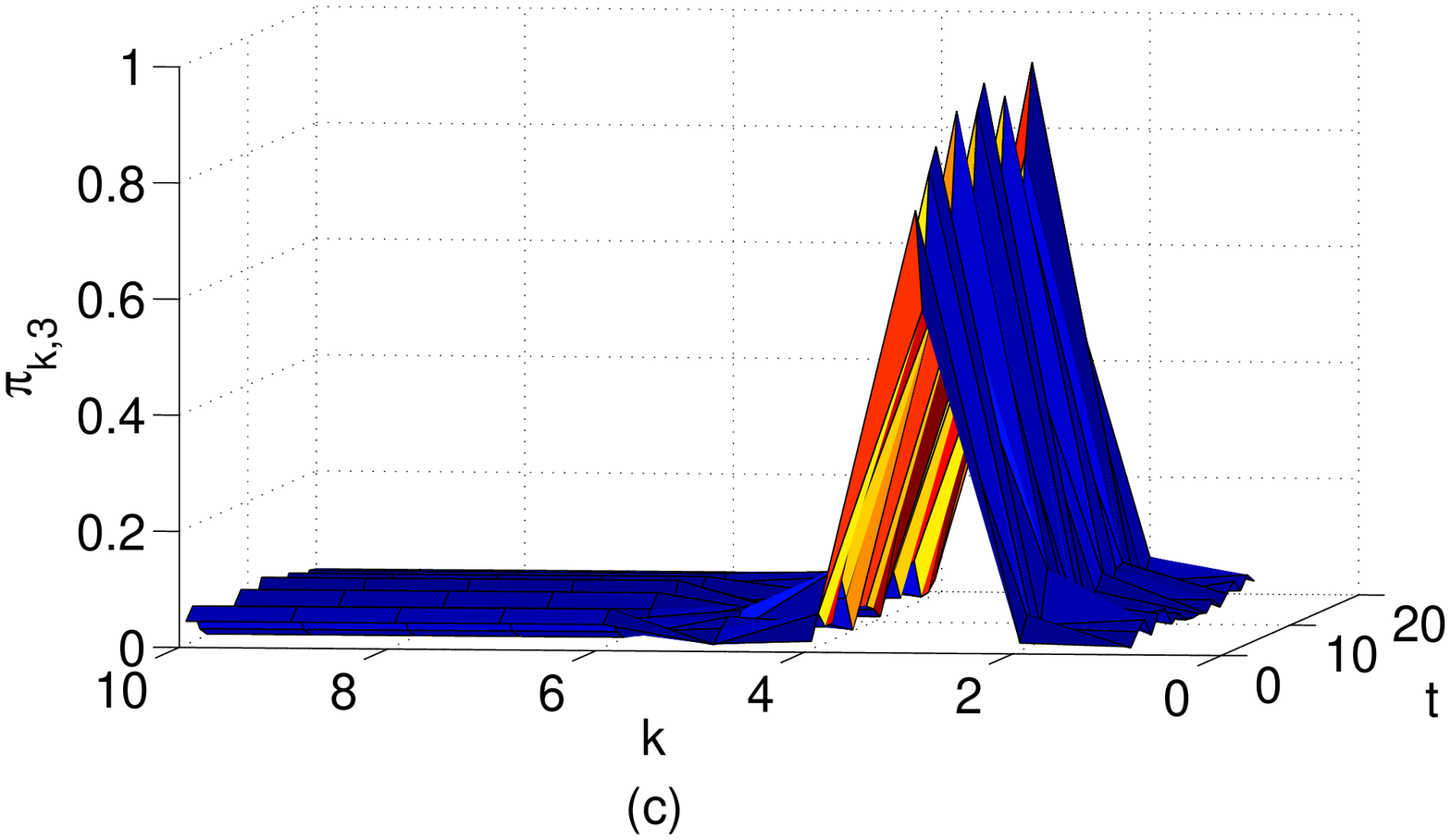}\vspace{0.1cm}\hspace{10cm}\includegraphics{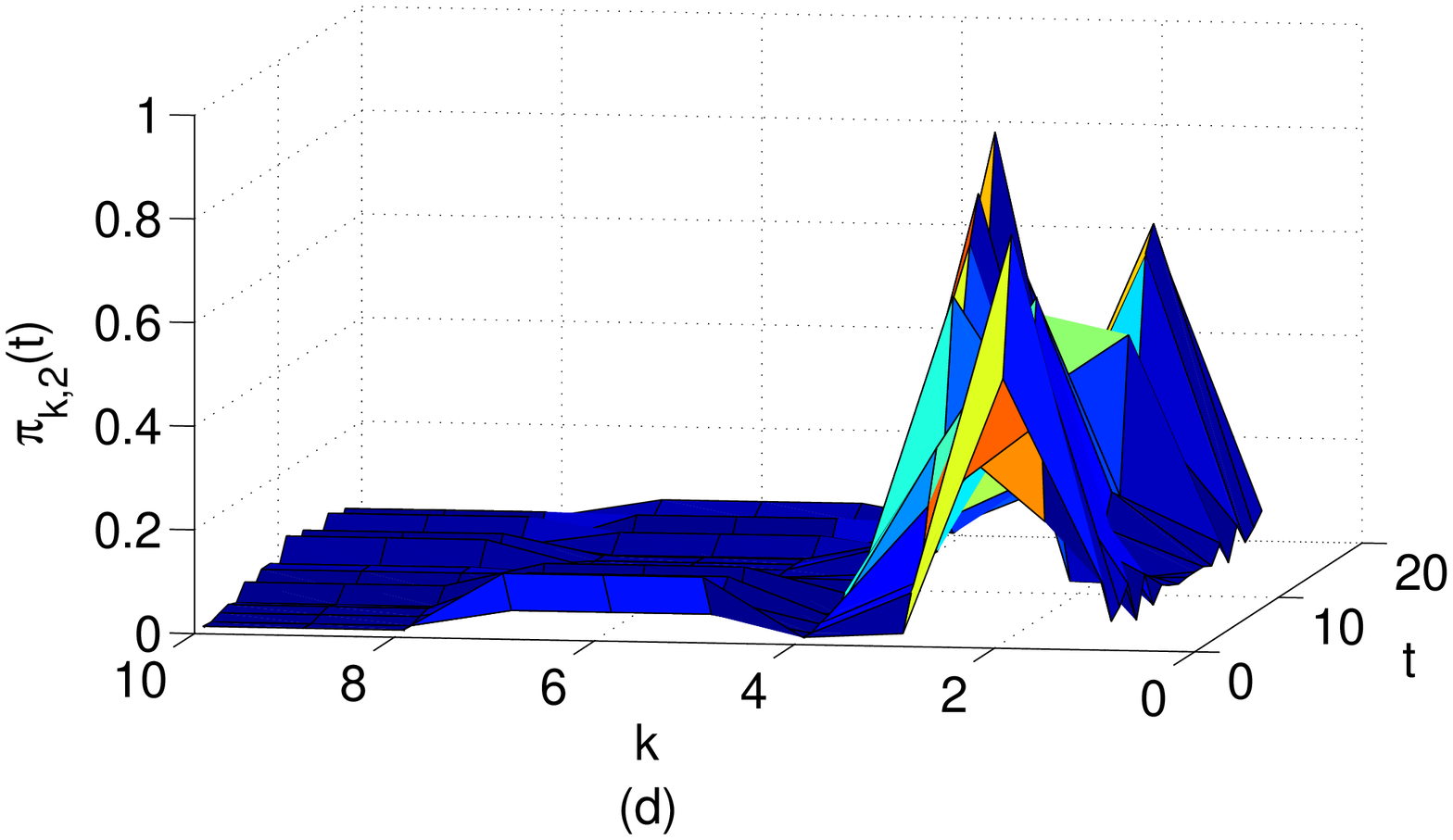}
\end{figure}

\begin{figure}[h]
\vspace{5cm}\hspace{0cm}\includegraphics{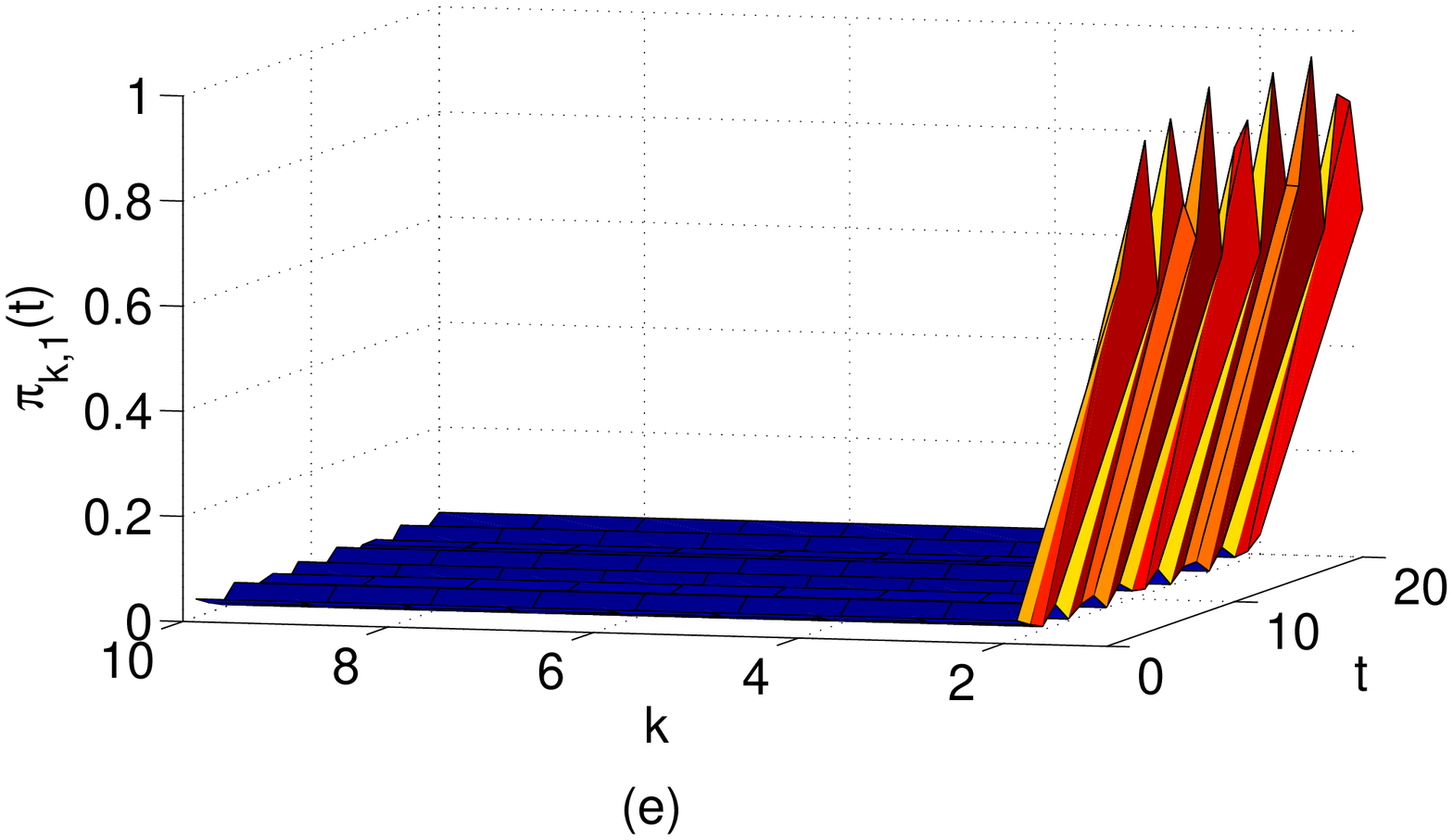}\vspace{-3cm}\caption{Figs. 2(a-e)
show the quantum transition probabilities on networks(a-e).}
\end{figure}

\begin{figure}[h]
\vspace{6cm}\hspace{-4cm}\includegraphics{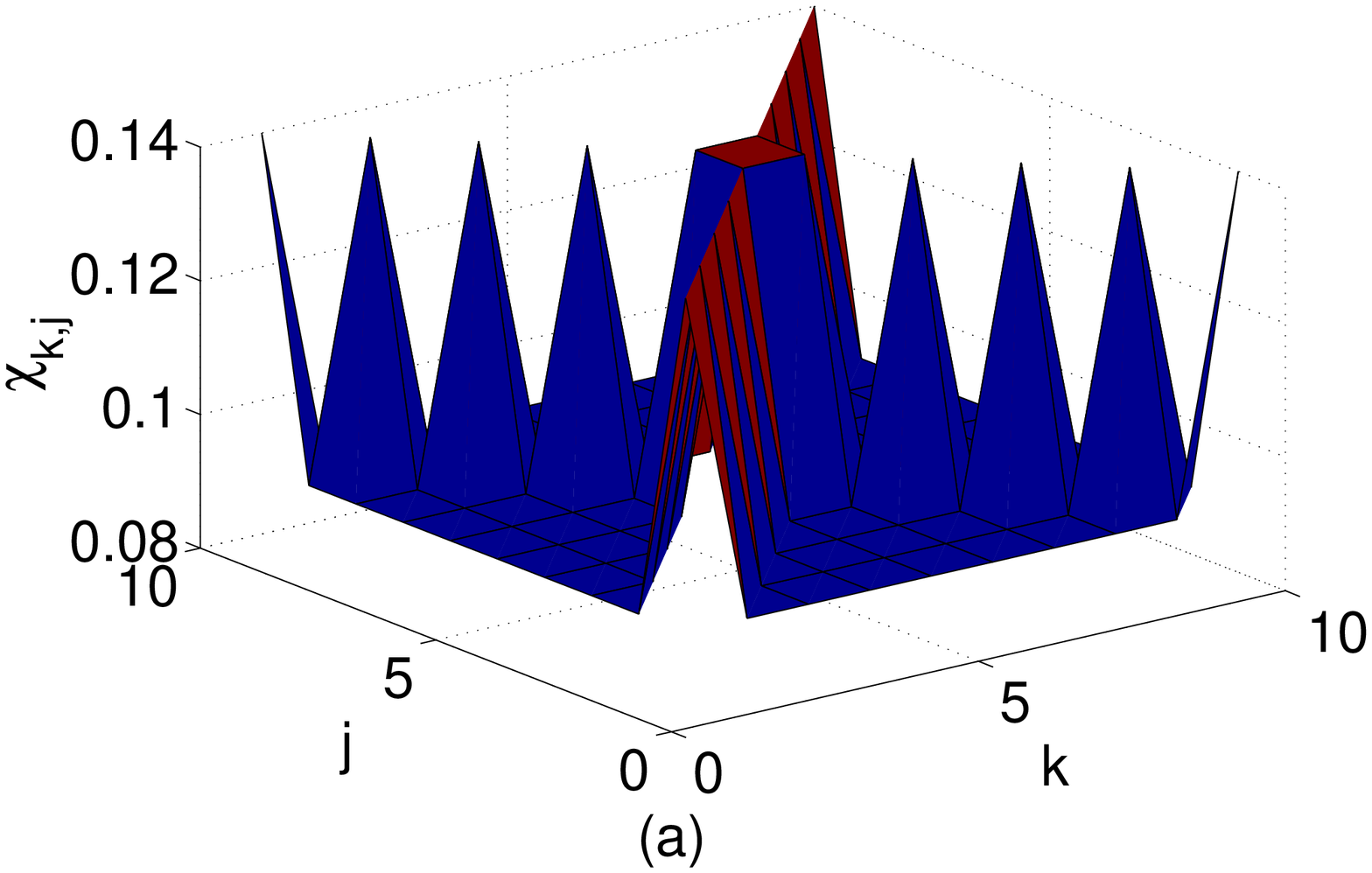}\vspace{-0.1cm}\hspace{10cm}
\includegraphics{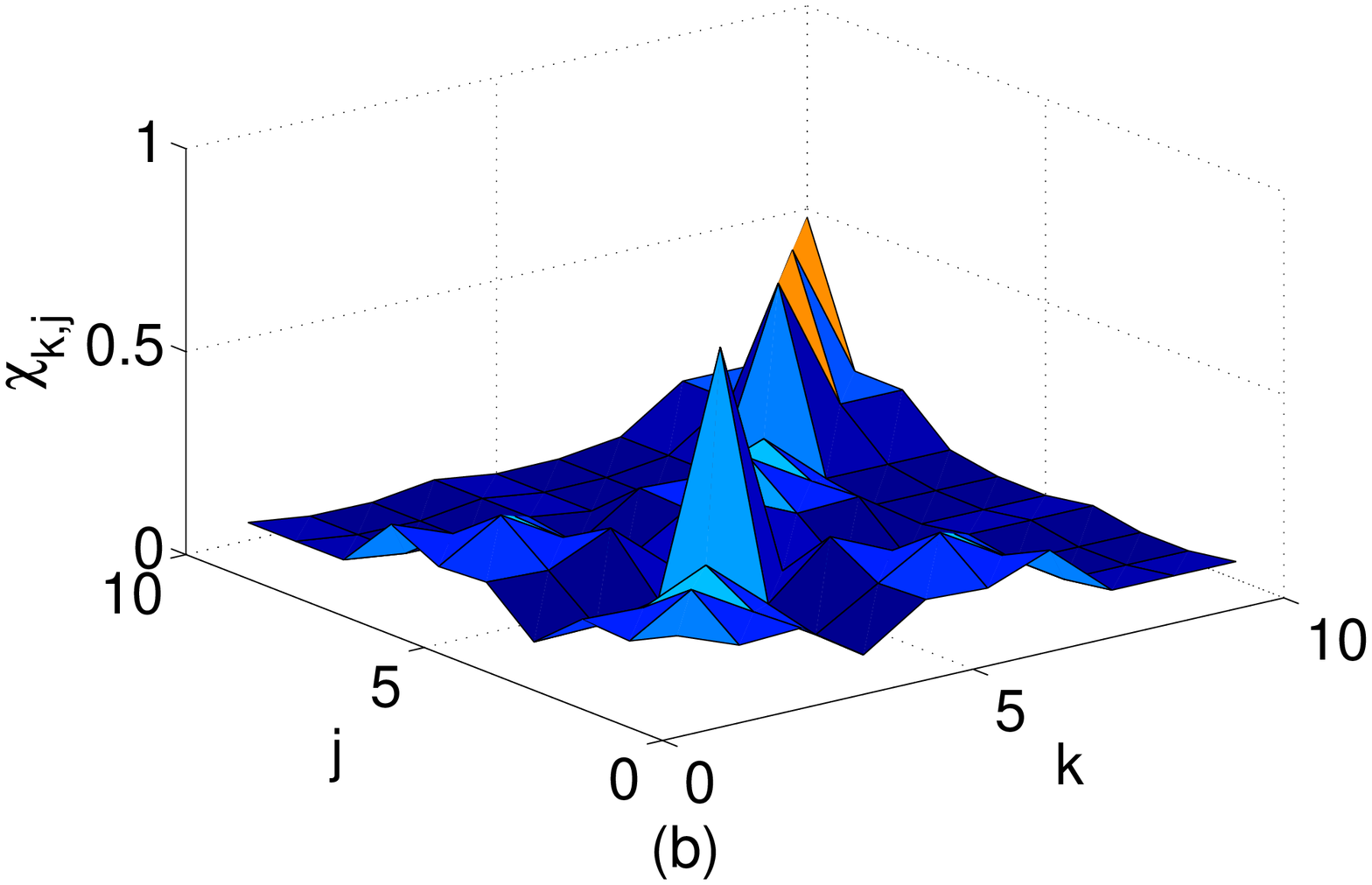}
\end{figure}

\begin{figure}[h]
\vspace{6cm}\hspace{-4.5cm}\includegraphics{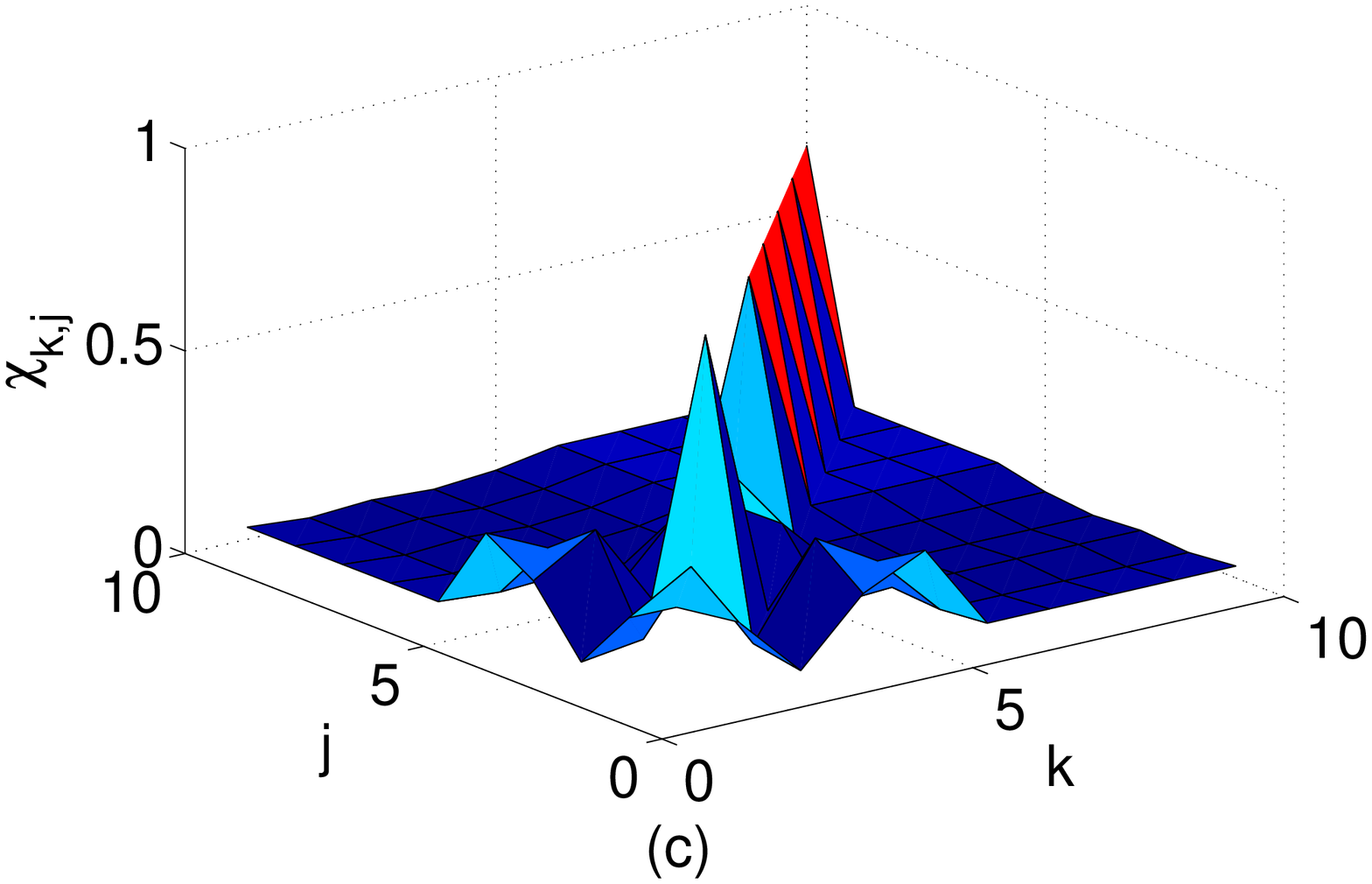}\vspace{0.1cm}\hspace{10.5cm}\includegraphics{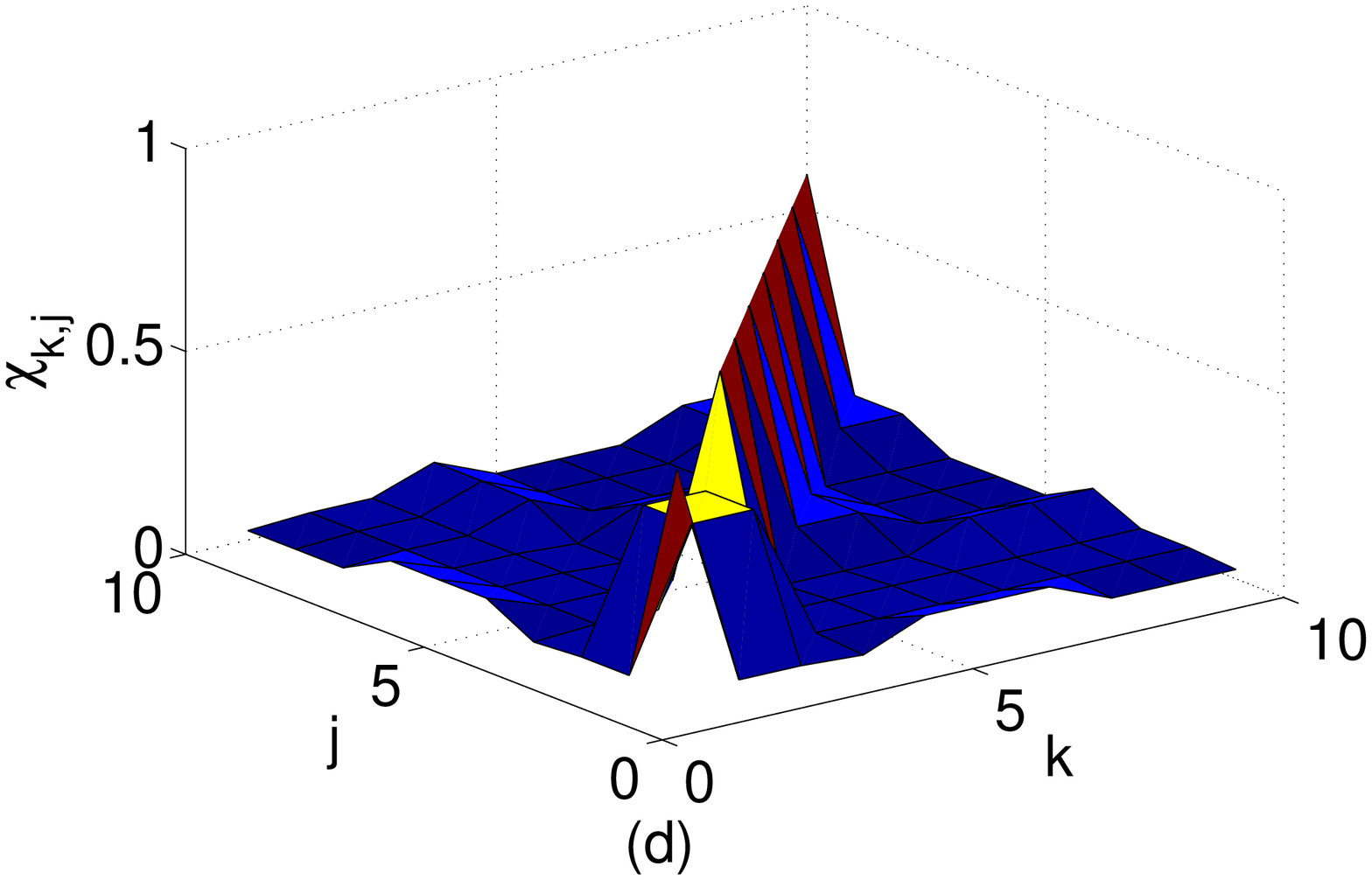}
\end{figure}

\begin{figure}[h]
\vspace{4cm}\hspace{0cm}\includegraphics{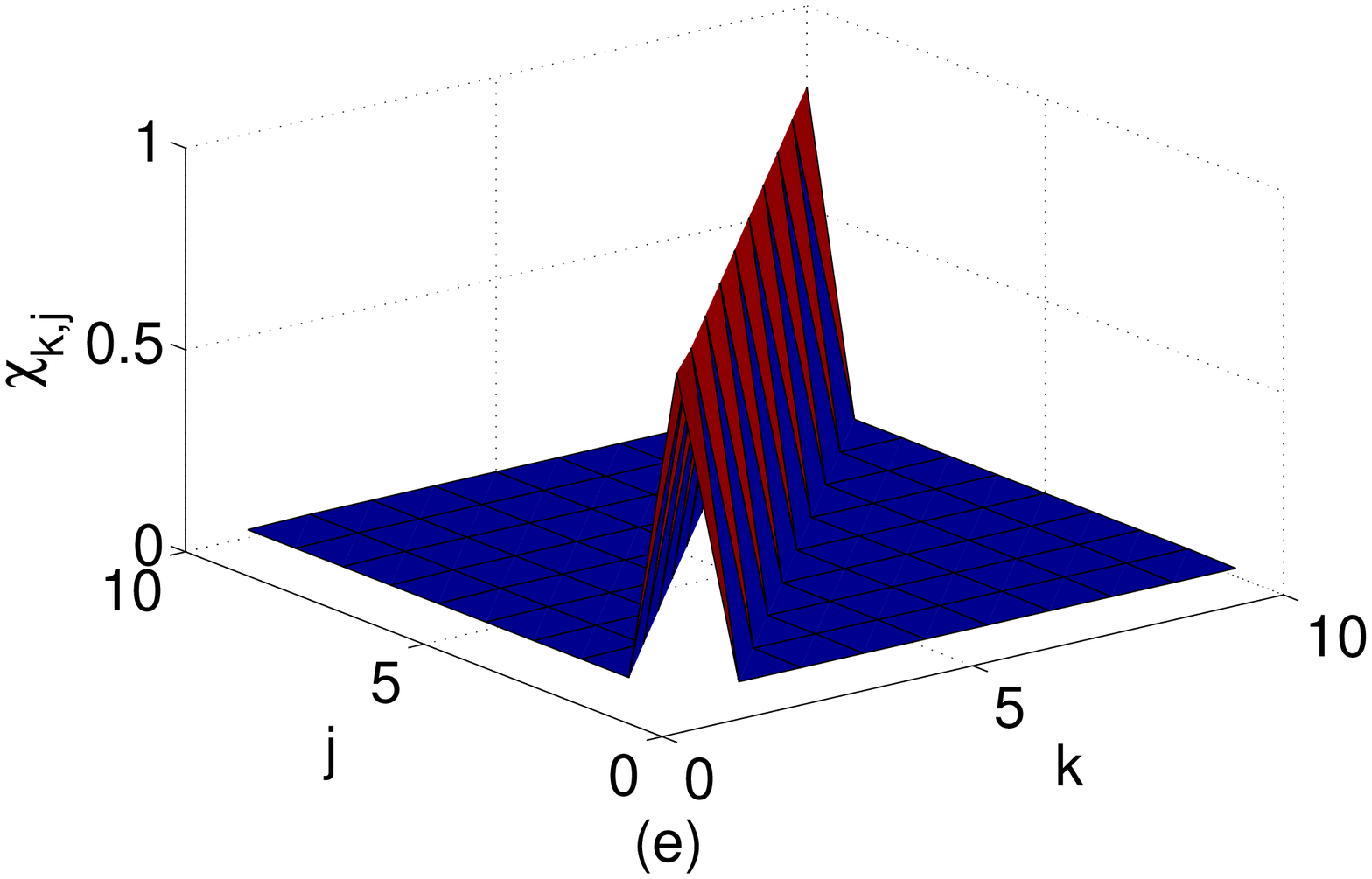}\vspace{-3cm}\caption{Figs. 3(a-e)
show limiting time average of quantum transition probabilities for
networks(a-e).}
\end{figure}
\clearpage  In Figs. 3(a-e), $\chi_{5,5}, \chi_{4,4}, \chi_{3,3},
\chi_{2,2}$ and $\chi_{1,1}$ have the high peaks, respectively.
These peaks are a consequence of the constructive interference due
to reflections at peripheral sites and boundaries. They prove that,
even at long times, the quantum probabilities depend on
significantly to the starting node which is consistent with findings
reported in~\cite{konno2, VMB, OMV} for Cayley tree(CT) and square
torus(ST). Hence, to get global information(independent of the
initial node) about CTQWs on networks(a-e), we must average over the
whole network nodes.
\section{Efficiency of classical and quantum walks}
The transport processes implemented by CTRWs and CTQWs will be more
efficient if the walker rapidly spreads over the network i.e. there
exists a fast delocalization~\cite{EAO}. On the other hand, a fast
delocalization implies a small probability to return(or stay) at the
initial node, thus the return probability can be good candidate to
evaluation the transport efficiency. But according the result of
Sec. 3, to get a global information about the transport efficiency,
we need a global quantity which is independent of the initial node.
For this aim, we average the return probability over all nodes of
network and obtain the average return probability.
\\Hence, the classical and quantum average return probabilities are calculated by the following simple expressions,
respectively~\cite{OMBE}
\begin{eqnarray}\label{7}
\bar{P}(t)=\frac{1}{N}\sum_{j=1}^{N}P_{j,j}(t)=\frac{1}{N}\sum_{n=1}^{N}e^{-E_{n}t}
\end{eqnarray}

\begin{eqnarray}\label{8}
\bar{\pi}(t)=\frac{1}{N}\sum_{j=1}^{N}\pi_{j,j}(t)=\frac{1}{N}\sum_{m,n,j=1}^{N}e^{-i(E_{n}-E_{m})t}|\langle
j|q_{n}\rangle|^{2}|\langle j|q_{m}\rangle|^{2}
\end{eqnarray}

It is evident that $\bar{P}(t)$ depends only on  the eigenvalues and
not to the eigenvectors of $T$ while $\bar{\pi}(t)$ depends on both
the eigenvalues and the eigenvectors of $H$. In the quantum case, we
can define a lower bound for $\bar{\pi}(t)$ which depends  only on
the $E_{n}$, i.e,:
\begin{eqnarray}\label{9}
\bar{\pi}(t)\geq\mid\bar{\alpha}(t)\mid^{2}=\mid\frac{1}{N}\sum_{n=1}^{N}e^{-iE_{n}t}\mid^{2}
\end{eqnarray}
where $\bar{\alpha}(t)=\frac{1}{N}\sum_{j=1}^{N}\alpha_{j,j}(t)$.
\\For the CTQW on a simple network with periodic boundary conditions,
the quantum average return probability equals to its lower bound
i.e. $\bar{\pi}(t)=|\bar{\alpha}(t)|^{2}$. The problem of the CTQW
on hypercube lattices in higher d-dimensional spaces separates in
every direction, thus we have
$\alpha_{j,j}^{(d)}=[\alpha_{j,j}(t)]^d$ which results in
$\bar{\pi}(t)=|\bar{\alpha}(t)|^{2}$, too~\cite{ABVBOM}. \\We denote
the degeneracy of eigenvalue $E_{n}$ by $D_{n}$ and rewrite Eqs.
(8), (10) as

\begin{eqnarray}\label{10}
\bar{P}(t)=\frac{1}{N}\sum_{E_{n}}D_{n}e^{-E_{n}t}
\end{eqnarray}

\begin{eqnarray}\label{11}
\bar{\pi}(t)\geq|\bar{\alpha}(t)|^{2}=\frac{1}{N^{2}}\sum_{E_{n},E_{m}}D_{n}D_{m}e^{-i(E_{n}-E_{m})t}.
\end{eqnarray}
According to Sec. 2, finally $\bar{\pi}(t)$ fluctuates about a
stationary(asymptotic) value given by $\bar{\chi}$:
\begin{eqnarray}\label{13}
\bar{\chi}=\lim_{T\longrightarrow\infty}\frac{1}{T}\int^{T}_{0}\bar{\pi}(t)dt
=\frac{1}{N}\sum_{n,m,j}\delta_{E_{n},E_{m}}|\langle
j|q_{n}\rangle|^{2}|\langle j|q_{m}\rangle|^{2}.
\end{eqnarray}
By Cauchy-Schwarz inequality or by taking the LTA of
$|\bar{\alpha}(t)|^{2}$, we can obtain a lower bound of $\bar{\chi}$
which does not depend on the eigenvectors~\cite{OMI}:

\begin{eqnarray}\label{14}
\bar{\chi}_{l b}=\frac{1}{N^{2}}\sum_{n,m}\delta_{E_{n},E_{m}}.
\end{eqnarray}
In fact this equation provides the exact asymptotic value of
$|\bar{\alpha}(t)|^{2}$ and the lower bound of asymptotic value of
$\bar{\pi}(t)$. Since some eigenvalues of $H$ might be degenerate,
the above sum is equal to the number of non-degenerate eigenvalues
plus the number squared of degenerate eigenvalues.
\\The relation among the average return probability and the
efficiency of transport can be considered from two different points
of view. From  the first point of view, the decay rate of the
average return probability is proportional to the transport
efficiency. The reason being that a quick decrease of the average
return probability results -on average- in a quick increase of the
probability for the walker to be at any other but the initial node,
which provides a more efficient transport. From the second point of
view, the asymptotic value of the average return probability has a
inverse relation with the transport efficiency, because a large
asymptotic value of the average return probability implies a large
probability to return at the initial node, which means inefficiency.
\\In~\cite{OMI}, authors studied the decay rate of $\bar{P}(t)$ and
$|\bar{\alpha}(t)|^{2}$ on the networks with two distinct eigenvalue
spectrums: uniform degeneracy and one highly degeneracy.
\\They numerically show that for the large
networks($N\gg 1$) whose eigenvalues have uniform degeneracy, the
quantum walk can be more efficient than the classical one. Also, for
the networks whose only eigenvalue $E_{l}$ has high degeneracy
$D_{l}$, they used the following approximate equation
\begin{eqnarray}\label{16}
\begin{array}{cc}
  |\bar{\alpha}(t)|^{2} & \approx\displaystyle\frac{1}{N^{2}}[D_{l}^{2}+2\displaystyle\sum_{E_{n}\neq
E_{l}}D_{n}D_{l}\cos((E_{n}-E_{l})t)].
\end{array}
\end{eqnarray}
By this equation, they found that $|\bar{\alpha}(t)|^{2}$ does not
show a decay to values  fluctuating about $1/N$ but rather to values
fluctuating about $1-1/N$, thus the quantum transport is less
efficient.
\\
\\In the following, we study
the classical and quantum efficiency on the small networks(with few
nodes) mentioned in Figs. 1(a-e). Note that all networks(a-e)
consist of 10 nodes and 9 links. The symmetry of network is given by
whose nodes having the similar situations. For example, nodes 8,9,10
in network(b) and nodes 5,6,7,8,9,10 in network(d) have the similar
situations, thus network(d) is  symmetrically higher than
network(b). On the other hand, the numerical determination of the
eigenvalues of networks(a-e) indicates that network(a) has no
degenerate eigenvalue while networks(b-e) have one degenerate
eigenvalue 1 with the order of degeneracy 2, 4, 6 and 8,
respectively. Hence, in networks(a-e) the increasing of network
symmetry results in the increasing of degeneracy of eigenvalue 1. We
denote the degeneracy of eigenvalue 1 by $D_{l}$ and represent it as
the degree of network symmetry. We divide the problem into two
separate cases as follows:
\subsection{non-degenerate eigenvalues}
Here, we consider network(a) with non-degenerate eigenvalue
spectrums. For this network, Eq. (12) can be written as

\begin{eqnarray}\label{15}\nonumber
  |\bar{\alpha}(t)|^{2} =\frac{1}{N^{2}}\displaystyle\sum_{E_{n},E_{m}}e^{-i(E_{n}-E_{m})t}.
\end{eqnarray}

From the above equation, one can  infer that after long time the
only terms with $E_{n}=E_{m}$ contribute to the sum and therefore,
the equation becomes of order $O(\frac{1}{N})$.
\\Fig. 4(a) shows the temporal behavior
of $\bar{P}(t)$, $\bar{\pi}(t)$ and $|\bar{\alpha}(t)|^{2}$ for
network(a). We can see that the classical curve(blue line) does not
show the constant value at the intermediate times whereas after
$t\approx N=10s$, not only $|\bar{\alpha}(t)|^{2}$(solid red line)
but also $\bar{\pi}(t)$(green line) fluctuate about the
equipartition value $1/10$. As mentioned above, the exact asymptotic
value of $|\bar{\alpha}(t)|^{2}$ and the asymptotic value of
$\bar{\pi}(t)$ can be reproduced by Eq. (14). Since the eigenvalue
spectrum of network(a) has no degenerate eigenvalue, the only the
number of non-degenerate eigenvalues contributes in Eq. (14),
resulting in the value $1/10$. To determine the scaling behavior of
$\bar{P}(t)$ and $|\bar{\alpha}(t)|^{2}$, we use the dashed
block($t^{-1/2}$) and red($t^{-1}$) lines, respectively.
 These lines show that the quantum return probability
decreases faster than classical one, thus the quantum walk is more
efficient than the classical random walk.

\subsection{degenerate eigenvalues}
In the following, we consider the transport efficiency on
networks(b-e) with degenerate eigenvalue spectrums. Figs. 4(b-e)
represent the temporal behavior of $\bar{P}(t)$,
$|\bar{\alpha}(t)|^{2}$ and $\bar{\pi}(t)$ for networks (b-e),
respectively.

 In all the figures, for all times, one can see that the
quantum average return probability(green curve) is higher than the
corresponding classical probability(blue curve), and thus the
quantum transport is less efficient than the classical one. \\On the
other hand, from Figs. 4(b-e), we find that after some time the
strong maxima of $\bar{\pi}(t)$ on highly symmetry networks(d,e) can
well be reproduced by the lower bound $|\bar{\alpha}(t)|^{2}$. Also,
the increasing of symmetry causes the dashed block curve of Eq.
(15), which is plotted for $E_{l}=1$, to become very close to the
full expression for $|\bar{\alpha}(t)|^{2}$. For instance, in Fig.
4(e) the positions of the exact curve of $|\bar{\alpha}(t)|^{2}$ and
its approximate equation almost coincide. It means that the only
highly degenerate eigenvalues contribute to $|\bar{\alpha}(t)|^{2}$,
and  there are only slight deviations due to the remaining ones.
Therefore, for highly symmetrical networks(d,e), we can well apply
Eq. (15) which proves that, always, $|\bar{\alpha}(t)|^{2}$
fluctuate about $1-1/N$ which is larger than the classical
equipartition $1/N$. Thus, the classical transport is more efficient
than the quantum one which is in agreement with the above result.
\begin{figure}[h]
\vspace{6cm}\hspace{-4cm}\includegraphics{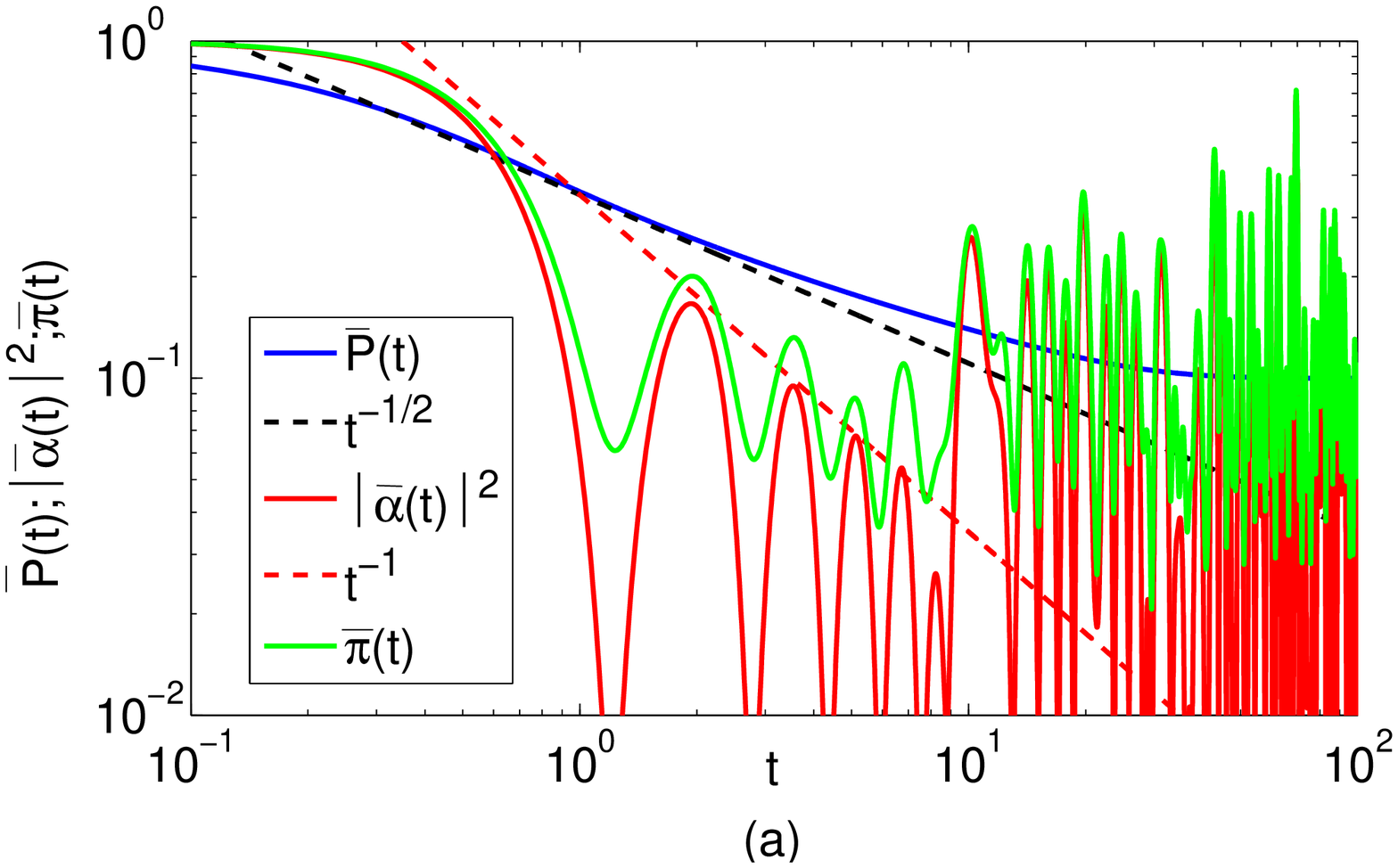}\vspace{-0.1cm}\hspace{10cm}
\includegraphics{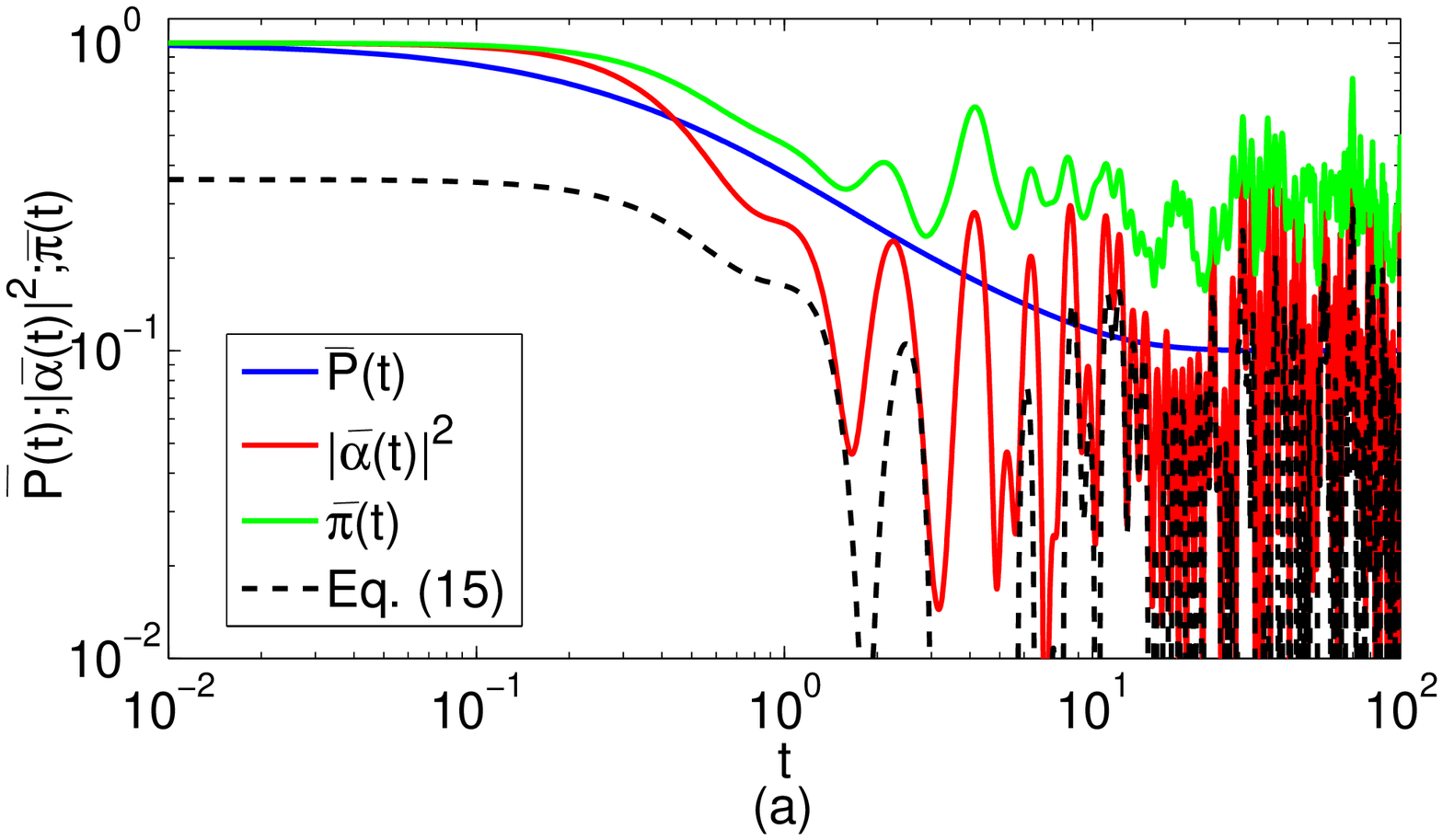}
\end{figure}

\begin{figure}[h]
\vspace{5.2cm}\hspace{-3.8cm}\includegraphics{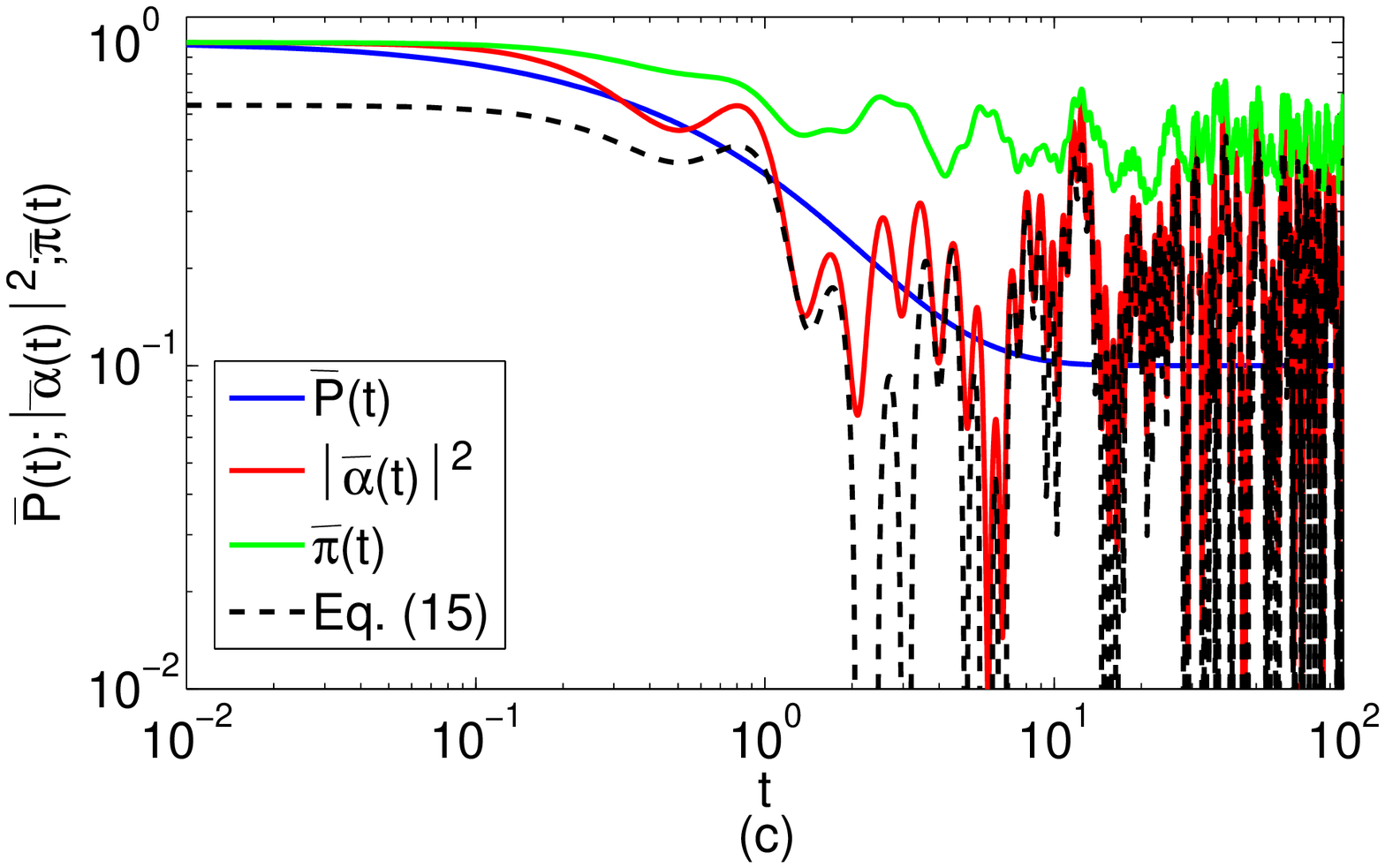}\vspace{0.15cm}\hspace{10cm}\includegraphics{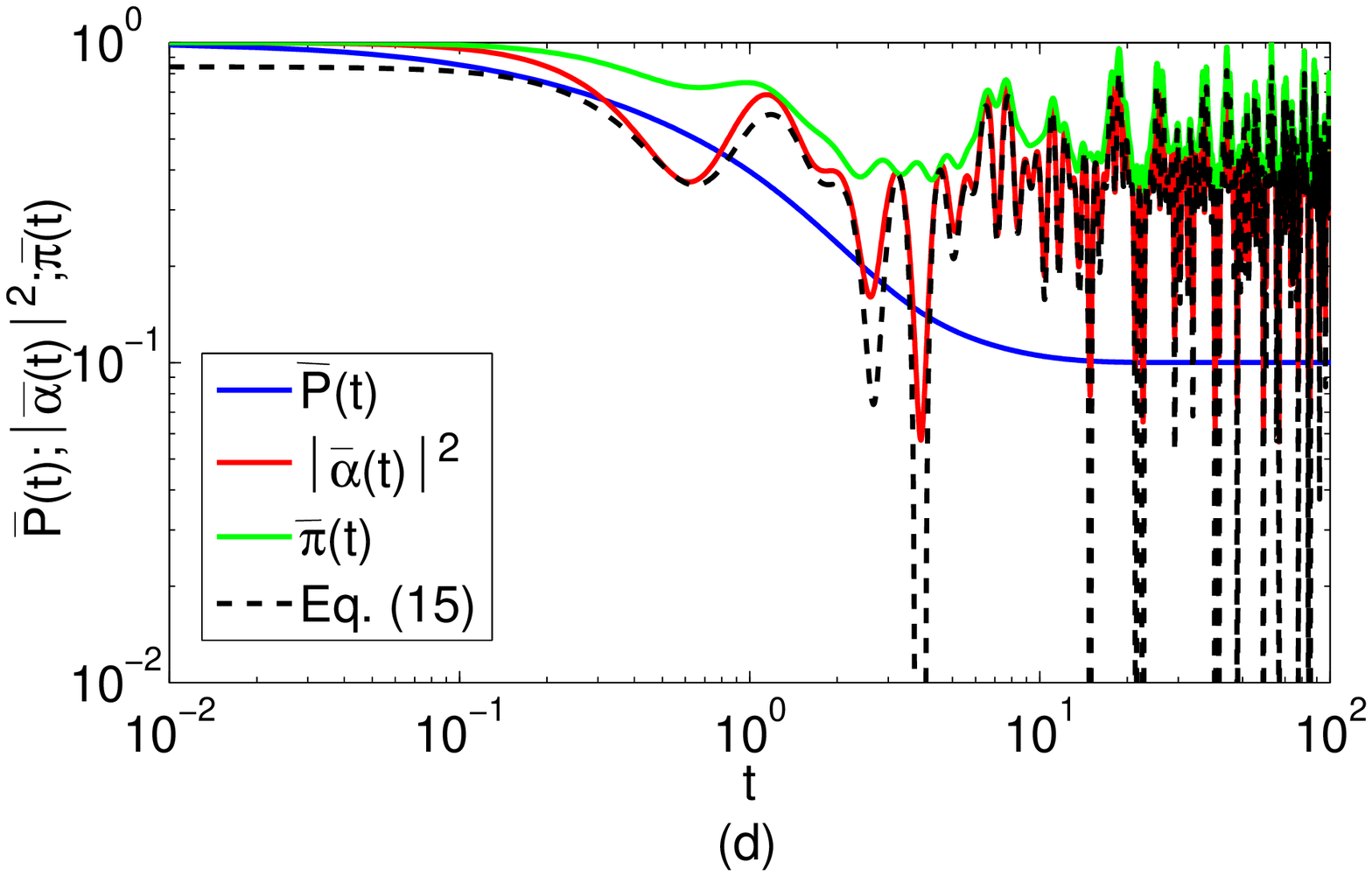}
\end{figure}

\begin{figure}[h]
\vspace{5cm}\hspace{0cm}\includegraphics{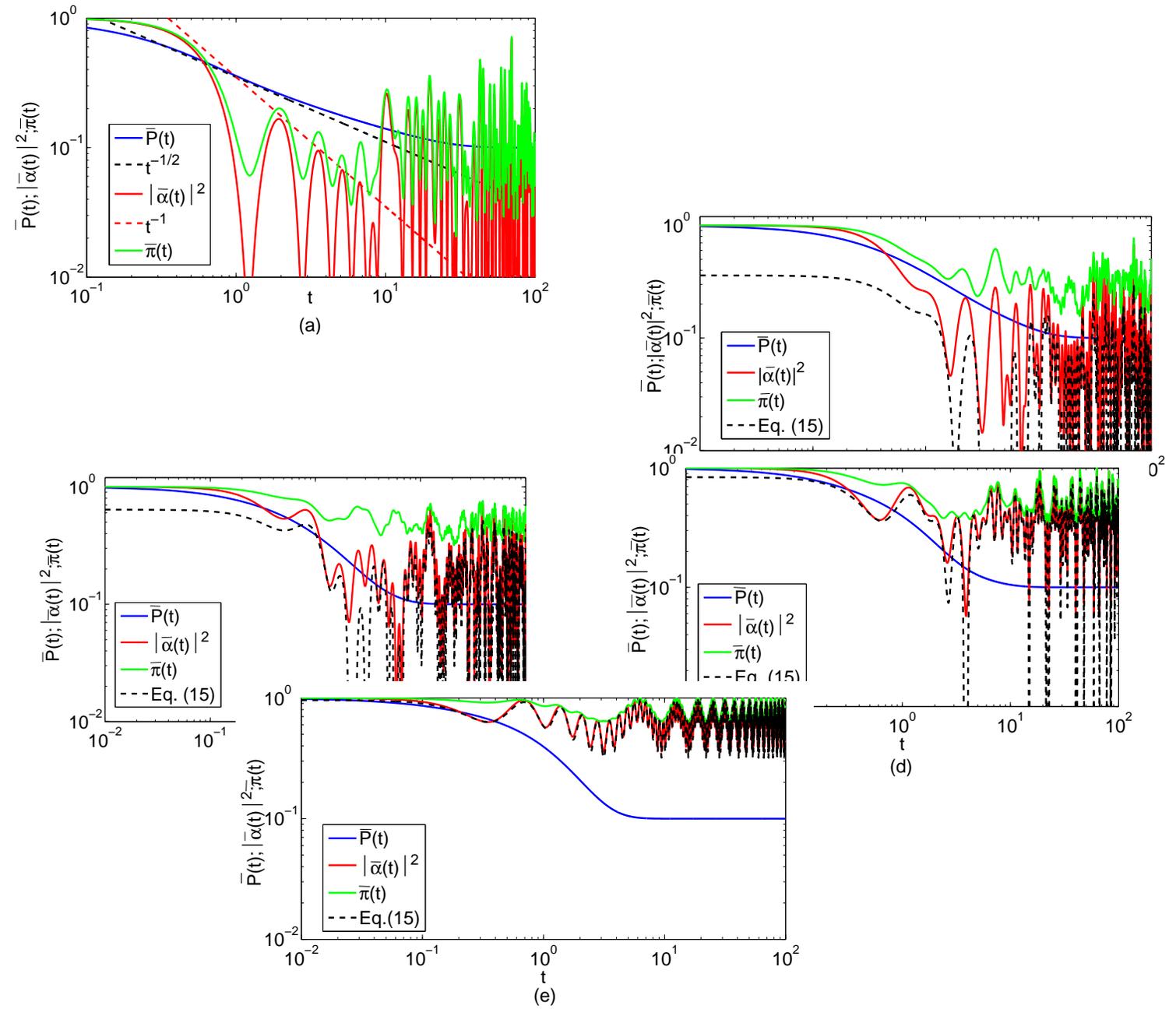}\vspace{-3cm}\caption{Figs. 4(a-e)
show $\bar{P}(t)$, $\bar{\pi}(t)$, $|\bar{\alpha}(t)|^{2}$ and its
approximate value for networks(a-e), respectively.}
\end{figure}
\clearpage Now, to study the details of the effect of symmetry in
the efficiency of classical and quantum transport, we study the
asymptotic values of $\bar{P}(t)$ and $|\bar{\alpha}(t)|^{2}$.
\\Network(b) have eigenvalue 1 with degeneracy 2-fold, (i.e.
$D_{l}=2$). In Fig. 4(b), after $t\sim 30$, we observe that
$\bar{P}(t)$ reaches equipartition 0.1, while
$|\bar{\alpha}(t)|^{2}$ fluctuates about the asymptotic value
\textbf{$0.12$}. Since the eigenvalue spectrum of network(b)
includes eigenvalue 1 with degeneracy 2 and eight non-degenerate
eigenvalues, Eq. (14) gives the asymptotic value of
$|\bar{\alpha}(t)|^{2}$ as $0.12$ which confirms the numerical
result.
\\In network(c), the degeneracy of eigenvalue 1 is 4-fold and the other eigenvalues are non-degenerate.
In Fig. 4(c), we see that at about $t\sim 12s$ the classical curve
reaches the equipartition probability $0.1$ while the lower bound
$|\bar{\alpha}(t)|^{2}$ fluctuates about the saturation value
\textbf{$0.22$}, as can be derived from Eq. (14).
\\Network(d) have eigenvalue 1 with degeneracy 6 and other eigenvalues non-degenerate.
In Fig. 4(d), one can see that the classical curve reaches $0.1$ at
$t\sim20s$, whereas the lower bound $|\bar{\alpha}(t)|^{2}$
fluctuates about the asymptotic value \textbf{$0.4$}(as can be
inferred of Eq. (14)).
\\Network(e) can well be represented a graph star with 10 nodes whose eigenvalues have three discrete values:
eigenvalue 1 with degeneracy 8, non-degenerate eigenvalues 0 and
10~\cite{OMBE}. Fig. 4(e) shows that after $t\sim 8s$, classical
probabilities reach $0.1$, while not only $|\bar{\alpha}(t)|^{2}$
but also $|\bar{\pi}(t)|^{2}$ fluctuate about a constant value
\textbf{$0.66$}(as can be obtained from Eq. (14)).
\\
We can conclude that, on symmetrical small networks such as those
analyzed here,
 CTRWs can spread faster than their quantum counterparts.
Moreover, the classical curves(blue lines) are flattened out and
$\bar{P}(t)$ tends towards a limiting value($\frac{1}{N}$) such that
with increasing the network symmetry this asymptotic domain is
obtained more quickly, which implies a more efficient classical
transport, except for networks(c),(d), i.e. however network(d) is
more symmetrical than network(c)($D_{l}$ of network(d)$>$$D_{l}$ of
network(c)), the limiting value for it occurs more late. \\But with
increasing the degree of network symmetry($D_{l}$), the quantum
average return probabilities fluctuate about a larger saturation
value, which implies a less efficient quantum transport, and there
is not any exception among networks(a-e).
\\The reason of these different behaviors can infer from Eqs. (8,9).
Based on Eq. (8), in the classical transport the eigenvalues of the
transfer matrix itself dominates the average return probability and
the degeneracy of eigenvalues directly dose not play role in
increase or decreasing $\bar{P}(t)$, while quantum
mechanically(Eq.(9)), the degeneracies of eigenvalues are governing
$\bar{\pi}(t)$. Therefore, as the above numerical results showed,
while the efficiency of classical transport on the mentioned
networks has not any exact relation with symmetry, increasing the
degree of symmetry decreases the efficiency of quantum transport on
them.
\section{Conclusions}
In summery, we studied CTRWs and CTQWs on some networks with few
nodes. We calculated the classical and quantum transition
probabilities on the networks and by numerical analysis found that
there is high probability to find the walker at the initial node for
the CTQW on the mentioned networks due to interference phenomenon,
even at long times. Thus, to get information about the transport
efficiency, we averaged  the quantity of the return probability over
the all nodes of network. Then, we studied the efficiency of
transport on the mentioned networks by studying the decay rate and
the asymptotic value of the average return probability. The
numerical results showed that the existence of symmetry in the
mentioned networks causes the quantum walks to be less efficient
than their classical counterparts. Moreover, we found that the
increasing of the symmetry of these networks decreases the
efficiency of quantum transport on them.

\end{document}